\documentclass[12pt]{article}
\usepackage{epigraph}
\usepackage{fancyhdr}
\usepackage{amsmath}
\usepackage{graphicx}
\usepackage{afterpage}
\usepackage[hyperfootnotes=false]{hyperref}
\usepackage{epsfig,epic,eepic,units}
\usepackage{hyperref}
\usepackage{url}
\usepackage{amsthm}
\usepackage{longtable}
\usepackage{mathrsfs}
\usepackage{multirow}
\usepackage{bigstrut}
\usepackage{amssymb}
\usepackage{graphicx}
\usepackage{caption}
\usepackage{textcomp}
\usepackage{setspace}
\usepackage[margin=1.25in]{geometry}
\usepackage [english]{babel}
\usepackage [autostyle, english = american]{csquotes}
\MakeOuterQuote{"}

\newtheorem{theorem}{Theorem}
\newtheorem{example}{Example}

\newtheorem{proposition}{Proposition}

\DeclareMathOperator*{\argmax}{arg\,max}

\hypersetup{colorlinks=true, citecolor=blue,urlcolor=blue}
\begin{document}
\begin{singlespacing}
\title{{\textbf{Nash Bargaining Over Margin Loans to Kelly Gamblers}}}
\author{\sc Alex Garivaltis\footnote{Assistant Professor, Department of Economics, School of Public and Global Affairs, College of Liberal Arts and Sciences, Northern Illinois University, 514 Zulauf Hall, DeKalb IL 60115.  E-mail:  \href{mailto:agarivaltis1@niu.edu}{\tt agarivaltis1@niu.edu}.  Homepage:  \url{http://garivaltis.com}.  ORCID iD:  0000-0003-0944-8517.}\\Northern Illinois University}
\maketitle
\abstract{I derive practical formulas for optimal arrangements between sophisticated stock market investors (continuous-time Kelly gamblers or, more generally, CRRA investors) and the brokers who lend them cash for leveraged bets on a high Sharpe asset (i.e. the market portfolio).  Rather than, say, the broker posting a monopoly price for margin loans, the gambler agrees to use a greater quantity of margin debt than he otherwise would in exchange for an interest rate that is lower than the broker would otherwise post.  The gambler thereby attains a higher asymptotic capital growth rate and the broker enjoys a greater rate of intermediation profit than would obtain under non-cooperation.  
\par
If the threat point represents a complete breakdown of negotiations (resulting in zero margin loans), then we get an elegant rule of thumb:  $r_L^*=\left(3/4\right)r+\left(1/4\right)\left(\nu-\sigma^2/2\right)$, where $r$ is the broker's cost of funds, $\nu$ is the compound-annual growth rate of the market index, and $\sigma$ is the annual volatility.  We show that, regardless of the particular threat point, the gambler will negotiate to size his bets as if he himself could borrow at the broker's call rate. 
}
\end{singlespacing}
\begin{equation}
\boxed{r_L^*=\frac{r}{2}+\frac{\mu'\Sigma^{-1}\mu-r^2\cdot\textbf{1}'\Sigma^{-1}\textbf{1}-2\left(\overline{\Gamma}-\overline{\pi}\right)}{4\left[\textbf{1}'\Sigma^{-1}(\mu-r\textbf{1})-1\right]}.}
\end{equation}
\titlepage
\newpage
\section{Introduction.}
Harry Markowitz' fundamental mean-variance theory (Markowitz 1952) of investment transforms the $n$-dimensional portfolio hyperplane into his trademark (two-dimensional) mean-variance plane.  Even better, the Markowitz bullet gives us a one-dimensional frontier of efficient portfolios that yield the greatest possible reward for any given level of risk.  And given the freedom to borrow and lend cash, we need only focus on a single fund of risky assets that gives the greatest reward per unit of risk;  the curvature of the Markowitz bullet is thereby replaced by the straightness of the capital market line.  But the theory ends there;  the practitioner receives no further prescription than to just borrow and buy as many shares of the tangency portfolio as is permitted by his particular appetite for risk. 
\par
This lack of guidance in one-dimensional gambling problems was acutely felt by card-counter Edward O. Thorp, who required an appropriate criterion for sizing his bets in certain favorable situations (Thorp 1966) that he found at the Nevada blackjack tables (cf. with his 2017 autobiography).  The correct answer (``Fortune's Formula,'' Poundstone 2010) is called the \textit{Kelly criterion}, after John Kelly (1956), a physicist at Bell Labs.  As was the custom in statistical communication theory, Kelly started with a simple example that turned out to be typical of the whole situation.  He considered a long sequence of independent bets on horse races whereby the gambler knows the (stationary) win probabilities to more precision than the posted (even) odds.  This illuminating environment led him to formulate the concept of a fixed-fraction betting scheme, whereby the gambler bets the same fraction of his wealth on each race.  His famous criterion singles out the fixed-fraction betting scheme (or ``Kelly bet'') that generates the highest possible asymptotic per-bet capital growth rate.  In blackjack, say, the Kelly fraction is $b^*:=p-q$, where $p$ is the chance of winning the next hand and $q$ is the chance of losing.  For instance, if you have a $p:=50.5\%$ chance of winning the next hand, the criterion dictates that you should bet $1\%$ of your net worth at even odds.
\par
Kelly's theory extends easily to Black-Scholes (1973) markets whereby the asset price $S_t$ follows a geometric Brownian motion.  In this environment, the gambler ``bets'' the fixed fraction $b$ of his wealth on the stock over each differential time step $[t,t+dt].$  Apart from the fact that we now have a continuum of possible profit-and-loss outcomes
\begin{equation}
dS_t:=S_t\times\left(\mu\,dt+\sigma dW_t\right)
\end{equation}
\begin{equation}
d\left(\log S_t\right)=\left(\mu-\frac{\sigma^2}{2}\right)dt+\sigma\,dW_t,
\end{equation}the correct behavior is governed by substantially the same logic (cf. with Alex Garivaltis 2018,  Garivaltis 2019a-c, and Ordentlich and Cover 1998), since the random fluctuations $dW_t:=\epsilon\times\sqrt{dt}$ are identically (normally) distributed and independent across time.
\par
It is well known (cf. with Edward O. Thorp 2006) that the Kelly bet for this market is
\begin{equation}
b^*=\frac{\mu-r}{\sigma^2}=\frac{1}{2}+\frac{\nu-r}{\sigma^2},
\end{equation}where $\nu:=\mu-\sigma^2/2$ is the expected geometric growth rate of the asset price and $r$ is the interest rate at which the gambler can borrow and lend.  If we take $r:=2.44\%$ (which is the 1-month U.S. treasury yield as of this writing) along with some stylized parameters ($\nu:=0.09, \sigma:=0.15$) meant to represent the behavior of the S\&P 500 index, we get $b^*=3.42$.  That is, if the gambler himself had the opportunity to borrow at the risk-free rate, he would borrow $\$2.42$ for every dollar of his own equity. 
\par
The author presently borrows from Interactive Brokers at a rate of $3.9\%$ compounded annually\footnote{This was on March 12, 2019.  The rate has dropped since then, to 3.62\% as of August 13, 2019.}, or $3.83\%$ compounded continuously, which corresponds to $b=2.8$.  However, for a client that is borrowing more than $\$3$ million, the rate changes to $2.7\%$ compounded annually, or $2.66\%$ compounded continuously, with a corresponding Kelly bet of $3.32$.  All of this is illustrated in Table \ref{ibkr}, which gives Interactive Brokers' pricing schedule for U.S. dollar margin loans as of 3/12/2019.

\begin{table}[t]
  \begin{center}
    \caption{\sc Margin loan interest rates (for U.S. dollars) at Interactive Brokers, 3/12/2019.}
    \label{ibkr}
    \begin{tabular}{l|c|r|rr} 
      Tier & Interest Rate & $\log(1+\text{Interest Rate})$ & Kelly Bet, $b$\\
      
      \hline
      0-100,000 & 3.9\% & 3.83\% & 2.8\\
				100,000.01-1,000,000 & 3.4\% & 3.34\% & 3.01\\
      1,000,000.01-3,000,000 & 2.9\% & 2.86\%  & 3.23  \\
      3,000,000.01-200,000,000 & 2.7\% & 2.66\%   & 3.32 \\
    \end{tabular}
  \end{center}
\end{table}

\subsection{Contribution.}
Taking our inspiration from this broker-client relationship (which is both real and ongoing), we formulate and solve a Nash (1950) bargaining problem between a stock broker (that can borrow cash on the money market at the \textit{broker call rate}, $r$) and a continuous time Kelly gambler (or, more generally, a CRRA investor) to whom the broker issues margin loans at a marked-up interest rate $r_L>r$.  The present situation for U.S. retail consumers of margin loans is just this:  the broker posts a price $\overline{r}_L$ (presumably a monopoly price), and the Kelly gambler (i.e. the author) responds by demanding the corresponding growth-optimal quantity $\overline{q}$ of margin loans per dollar of account equity.  The corresponding Kelly bet is then $\overline{b}:=\overline{q}+1$.  These choices lead to a definite (logarithmic) capital growth rate of $\overline{\Gamma}$ for the gambler and a definite profit rate $\overline{\pi}$ for the broker.
\par
Obviously, the principals should cooperate, if possible, and negotiate a margin loan contract $\left(b^*,r_L^*\right)$ that jointly specifies the interest rate, the client's portfolio, and the quantity of margin loans to be issued over the differential time step $\left[t,t+dt\right]$.  It would thereby be possible to convert the deadweight loss of monopoly into some agreed upon surplus values $\Gamma^*-\overline{\Gamma}$ and $\pi^*-\overline{\pi}$.  We show that, regardless of the disagreement point $\left(\overline{\pi},\overline{\Gamma}\right)$, the client will negotiate to bet as if he himself had the opportunity to borrow at the broker's cost of funds.  We derive exact formulas for the efficient profit-growth frontier, the negotiated behavior $\left(b^*,r_L^*\right)$, and the final utilities $\left(\pi^*,\Gamma^*\right)$ that obtain from Nash's cooperation scheme.  We show that the efficient frontier is a straight line (whose slope is $-1$), and therefore our bargaining problem is one of transferable utility (cf. with Mas-Colell, Whinston, and Green 1995).  We find that the Nash bargaining solution given here coincides with the Egalitarian solution, whereby all surplus value gained from cooperation is shared equally;  the correct outcome $\left(\pi^*,\Gamma^*\right)$ on the profit-growth frontier is found by intersecting it with a $45$\textdegree\,line emanating from the threat point $\left(\overline{\pi},\overline{\Gamma}\right)$.

\section{Definitions and Notation.}
As indicated above, we consider a Black-Scholes (1973) market with a single risk asset (i.e. the S\&P 500 index) whose price $S_t$ follows the geometric Brownian motion
\begin{equation}
\frac{dS_t}{S_t}:=\mu\,dt+\sigma dW_t,
\end{equation}where $\mu$ is the annual drift rate, $\sigma$ is the volatility, and $W_t$ is a standard Brownian motion.  We consider a broker that makes margin loans at a continuously-compounded interest rate of $r_L$ per year, and whose cost of funds (``broker call rate'') is denoted by $r$.  The broker makes these loans to a continuous-time Kelly (1956) gambler (cf. with David Luenberger 1998) whose behavior is characterized by the fact that he continuously maintains some fixed level $b$ of exposure to the risk asset.  That is, the gambler continuously maintains the fraction $b$ of his wealth in the stock;  if $b>1$ then he continuously maintains a margin (debit) balance in the amount of $b-1$ of his wealth.  For instance, if $b:=1.5$ then the gambler's margin loan balance would be continuously adjusted so as to constitute $50\%$ of his wealth.  We let $V_t(b)$ denote the gambler's wealth process, where $V_0$ is some given initial wealth.  The (dollar) quantity of margin loans demanded is $q:=\left(b-1\right)V_t(b)$.  The broker's instantaneous rate of profit per year is $\pi\left(b,r_L\right):=\left(r_L-r\right)q=\left(r_L-r\right)\left(b-1\right)V_t(b)$, where $r_L-r$ is the net interest margin.
\par
We will assume that the client's objective is to maximize the almost-sure continuously-compounded asymptotic growth rate of his capital, i.e. $\Gamma:=\lim\limits_{t\to\infty}\left(1/t\right)\log\left[V_t(b)/V_0\right]$.  As we will see presently (cf. with David Luenberger 1998), this is equivalent to maximizing the drift of $\log V_t(b)$.  The gambler's fortune evolves according to
\begin{equation}
\frac{dV_t(b)}{V_t(b)}=b\frac{dS_t}{S_t}-(b-1)r_L\,dt=\left[r_L+(\mu-r_L)b\right]dt+b\sigma\,dW_t.
\end{equation}
Since the gambler's fortune follows a geometric Brownian motion, one can apply It\^{o}'s Lemma (cf. with Paul Wilmott 2001) to obtain the relation
\begin{equation}
d\left(\log V_t(b)\right)=\left[r_L+\left(\mu-r_L\right)b-\frac{\sigma^2}{2}b^2\right]dt+b\sigma\,dW_t,
\end{equation}or, equivalently,
\begin{equation}
V_t(b)=V_0\times\exp\left\{\left[r_L+b\left(\mu-r_L\right)-\frac{\sigma^2}{2}b^2\right]t+b\sigma W_t\right\}.
\end{equation}Accordingly, the gambler's \textbf{\textit{asymptotic growth rate}} will be denoted $\Gamma\left(b,r_L\right)$:
\begin{equation}
\boxed{\Gamma\left(b,r_L\right):=r_L+b\left(\mu-r_L\right)-\frac{\sigma^2}{2}b^2=\lim_{t\to\infty}\frac{\log\left[V_t(b)/V_0\right]}{t}=\frac{\mathbb{E}\left[d\left(\log V_t(b)\right)\right]}{dt}.}
\end{equation}
\section{Nash Bargaining.}
In what follows, we will assume that at every instant $t$, the broker and the Kelly gambler will Nash bargain (Nash 1950) over the margin loan arrangement $\left(b,r_L\right)$, which simultaneously specifies the quantity of margin loans $q=\left(b-1\right)V_t$ and the interest rate $r_L$ that will be charged by the broker over the differential time step $\left[t,t+dt\right]$.  We let $\left(\overline{\pi},\overline{\Gamma}\right):=\left(\pi\left(\overline{b},\overline{r}_L\right),\Gamma\left(\overline{b},\overline{r}_L\right)\right)$ denote the \textit{threat point}, meaning that a breakdown in negotiations will lead to the broker charging $\overline{r}_L$ and to the gambler betting the fraction $\overline{b}$ of his wealth on the stock over $\left[t,t+dt\right]$.  Non-cooperation would thereby lead to a profit rate of $\overline{\pi}=\pi\left(\overline{b},\overline{r}_L\right)$ for the broker and a 
growth rate of $\overline{\Gamma}=\Gamma\left(\overline{b},\overline{r}_L\right)$ for the gambler.
\begin{example}\label{monopthreat}
Rather than cooperate, the broker posts a take-it-or-leave-it price $r_L\geq r$, and the client chooses the corresponding Kelly bet (or $\log$-optimal portfolio)
\begin{equation}
b^*\left(r_L\right):=\frac{\mu-r_L}{\sigma^2}=\argmax_{b\geq1}\,\,\Gamma\left(b,r_L\right),
\end{equation}where we must assume that $\mu-\sigma^2\geq r_L\geq r$ in order to guarantee that the client will take a margin loan.  This means that the risk asset must be sufficiently favorable (high drift and low volatility) in relation to the cost of funds, and also the interest rate $r_L$ must be sufficiently low that the client does not choose $b=1$.  The quantity of margin loans (per dollar of client equity) is
\begin{equation}
q=\frac{\mu-r_L}{\sigma^2}-1,
\end{equation}the broker's instantaneous rate of profit per unit time is
\begin{equation}
\overline{\pi}=\left(\frac{\mu-r_L}{\sigma^2}-1\right)\left(r_L-r\right),
\end{equation}and, after simplification, the gambler's asymptotic capital growth rate is
\begin{equation}
\overline{\Gamma}=r_L+\frac{1}{2}\left(\frac{\mu-r_L}{\sigma}\right)^2.
\end{equation}
\end{example}
\begin{example}\label{breakdown}
After a complete breakdown of negotiations, the broker does not even offer the client a margin loan (say, $r_L:=\infty$, or the client refuses to take a loan).  Thus, the client just buys and holds the stock ($b=1, q=0, \overline{\pi}=0$), achieving an asymptotic growth rate of $\overline{\Gamma}=\mu-\sigma^2/2$.
\end{example}Note that the total non-cooperation of Example \ref{breakdown} is not a credible threat (e.g. it is not subgame-perfect), because if the broker posts a reasonable price (as in Example \ref{monopthreat}), then the investor will optimally choose to borrow to the extent possible.
\par
Naturally, a negotiated contract has the potential to make both parties better off:  the client will agree to use more margin debt in exchange for a lower interest rate.  The contract will be arranged just so;  the gambler achieves a higher asymptotic growth rate on account of the lower $r_L$ and the broker achieves a higher rate of profit on account of the increased $b$.
\par
The Nash product (cf. with John Nash 1950) is equal to
\begin{equation}
N\left(b,r_L\right):=\left[\pi\left(b,r_L\right)-\overline{\pi}\right]\times\left[\Gamma\left(b,r_L\right)-\overline{\Gamma}\right],
\end{equation}and the Nash bargaining solution is equal to
\begin{equation}
\left(b^*,r_L^*\right):=\argmax_{\left\{b\geq1,\,\,r_L\geq r\right\}}\,\,N\left(b,r_L\right).
\end{equation}Note that, since $\pi\left(\bullet,\bullet\right)$ is directly proportional to $V_t(b)$, any Nash bargain that is struck for the duration of the differential time step $\left[t,t+dt\right]$ will be independent of the client's wealth level $V_t(b)$;  the negotiated behavior $\left(b^*,r_L^*\right)$ depends only on the GBM parameters $\left(\mu,\sigma\right)$, the broker's cost of funds ($r$), and the threat point $\left(\overline{\pi},\overline{\Gamma}\right)$\footnote{The principals are assumed to re-negotiate the margin loan contract after every differential tick $dt$ of the market clock.  However, the Nash bargaining solution never actually changes, on account of the fact that the Nash product is directly proportional to the client's wealth $V_t(b)$.}.  
\par
We should stress the fact that, although the client's welfare is measured in terms of his asymptotic almost sure capital growth rate (which is equal to his instantaneous expected compound-growth rate), the broker's welfare is measured by its instantaneous rate of intermediation profit per dollar of client equity.  Because this profit amounts to a fixed percentage of the client's wealth (e.g. the net interest margin times the client's debt-to-equity ratio), the asymptotic growth rate of the broker's fee income is in fact equal to client's long run capital growth rate, $\Gamma$.
\par
The first-order condition $\partial N/\partial r_L=0$ simplifies to
\begin{equation}\label{equal}
\Gamma\left(b,r_L\right)-\overline{\Gamma}=\pi\left(b,r_L\right)-\overline{\pi},
\end{equation}e.g. the asymptotic growth rate that the gambler gains from cooperation must be equal to the rate of profit (per dollar of client equity) that the broker gains relative to the threat point.  Taking the other first-order condition $\partial N/\partial b=0$ and using (\ref{equal}) to simplify, we get
\begin{proposition}
Under Nash bargaining, regardless of the threat point, the Kelly gambler will bet the fraction
\begin{equation}
\boxed{b^*=\frac{\mu-r}{\sigma^2}},
\end{equation}e.g. he will negotiate to bet as if he himself could borrow at the broker's call rate, $r$.
\end{proposition}Note that the negotiated bet size $b^*=\left(\mu-r\right)/\sigma^2$ amounts to a special case of the usual Merton strategy $b:=(\mu-r)/(\gamma\sigma^2)$ (cf. with Merton 1969 and Merton 1990) that obtains for an investor with CRRA utility 
\begin{equation}
u(x):=\begin{cases}
\left(x^{1-\gamma}-1\right)/\left(1-\gamma\right) & \text{if }\gamma>0\text{ and }\gamma\neq1\\
\log x & \text{if }\gamma=1,\\
\end{cases}
\end{equation}
where $\gamma\equiv-x\cdot u''(x)/u'(x)$ denotes the investor's (constant) coefficient of relative risk aversion.  Substituting this value of $b$ into (\ref{equal}), solving, and simplifying, we get
\begin{theorem}
The negotiated interest rate under Nash bargaining is given by the formula
\begin{equation}\label{negotiated}
\boxed{r_L^*=\frac{r}{2}+\frac{\mu^2-r^2-2\sigma^2\left(\overline{\Gamma}-\overline{\pi}\right)}{4\left(\mu-\sigma^2-r\right)}.}
\end{equation}

\end{theorem}

\begin{example}
Under total non-cooperation ($\overline{\pi}=0$ and $\overline{\Gamma}=\mu-\sigma^2/2$), after factoring and simplifying, we get
\begin{equation}
\boxed{r_L^*=\frac{\mu+3r-\sigma^2}{4}=\frac{3}{4}r+\frac{1}{4}\left(\nu-\frac{\sigma^2}{2}\right),}
\end{equation}where $\nu:=\mu-\sigma^2/2$ is the asymptotic growth rate of the asset price $S_t$.  The broker's rate of profit is
\begin{equation}
\pi=\left(\frac{\mu-\sigma^2-r}{2\sigma}\right)^2.
\end{equation}
\end{example}
\begin{example}
For the parameters $\sigma:=15\%, r:=3\%, \nu:=9\%, \mu:=\nu+\sigma^2/2$, the continuous-time Kelly rule is $b^*=3.17$.  The negotiated interest rate is $r_L^*=4.2\%$, the net interest margin is $1.2\%$, and the gambler achieves an asymptotic continuously-compounded growth rate of $\Gamma^*=12\%$ per year.  The broker earns instantaneous margin loan profits at a rate of $\pi^*=2.6\%$ of the client's equity per year.
\end{example}
\subsection{Utility Possibility Frontier.}
In this subsection, we derive the utility possibility frontier, e.g. we calculate the maximum possible growth rate $\Gamma$ that is (cooperatively) achievable for a given profit rate $\pi$.  Solving for $r_L$ in terms of $\pi$, we get
\begin{equation}
r_L=r+\frac{\pi}{b-1}.
\end{equation}Substituting this expression into the formula for $\Gamma$, we get, after simplification,
\begin{equation}\label{frontier}
\Gamma=r+\left(\mu-r\right)b-\frac{\sigma^2}{2}b^2-\pi.
\end{equation}Maximizing with respect to $b$, we obtain $b^*=\left(\mu-r\right)/\sigma^2$.  Substituting back into (\ref{frontier}) and simplifying, we get the (linear) equation of the efficient growth-profit frontier:
\begin{equation}\label{frontier}
\boxed{\Gamma=r+\frac{1}{2}\left(\frac{\mu-r}{\sigma}\right)^2-\pi.}
\end{equation}Juxtaposing (\ref{frontier}) with the egalitarian condition $\Gamma-\overline{\Gamma}=\pi-\overline{\pi}$, we obtain the fact that the final utilities under Nash bargaining are
\begin{equation}
\boxed{\pi^*=\frac{r-\left(\overline{\Gamma}-\overline{\pi}\right)}{2}+\frac{1}{4}\left(\frac{\mu-r}{\sigma}\right)^2}
\end{equation}and
\begin{equation}
\boxed{\Gamma^*=\frac{r+\overline{\Gamma}-\overline{\pi}}{2}+\frac{1}{4}\left(\frac{\mu-r}{\sigma}\right)^2.}
\end{equation}
\section{Monopoly Threat Point.}\label{monopoly}
\begin{figure}[t]
\begin{center}
\includegraphics[height=250px]{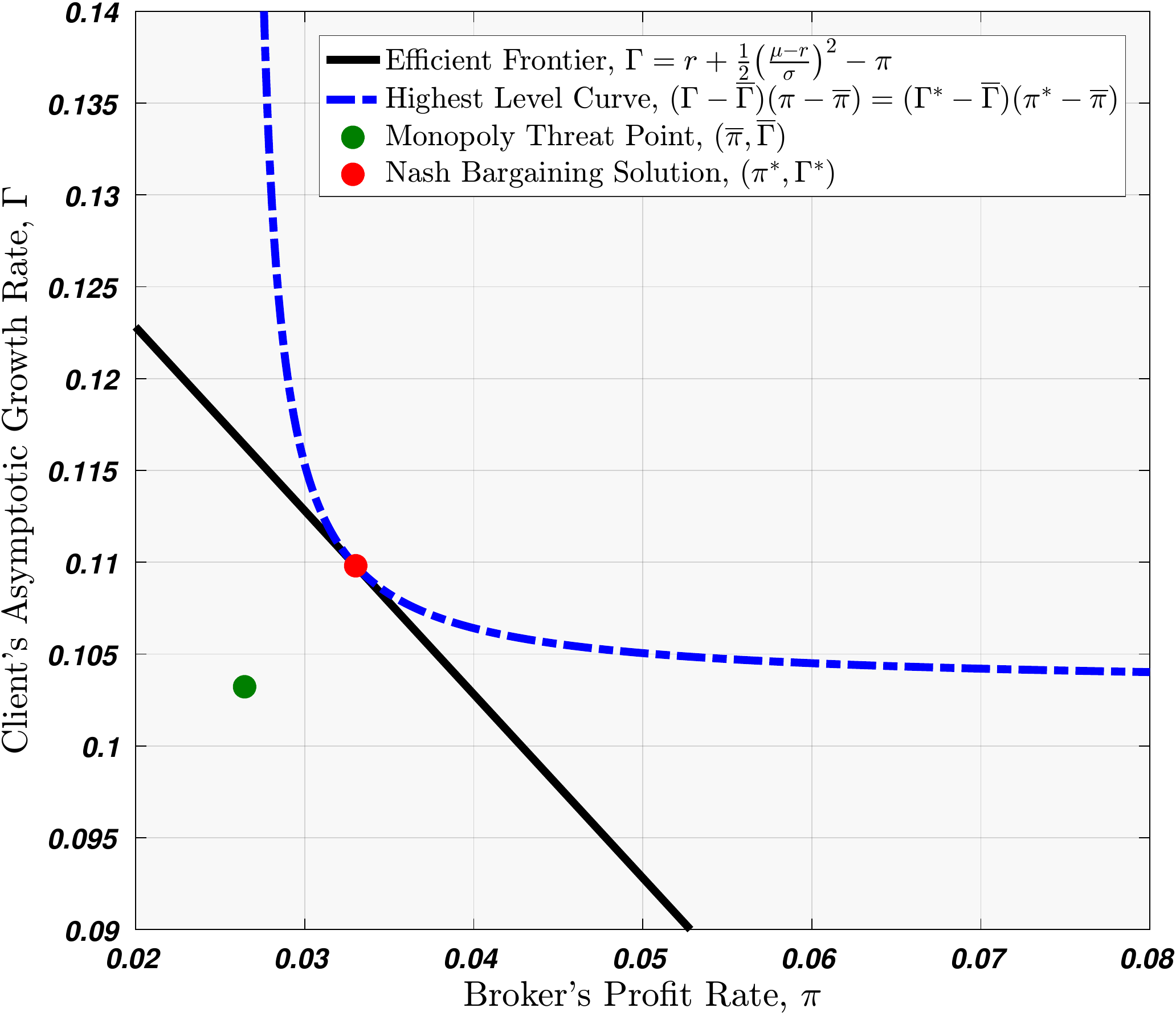}
\caption{\sc Gains from cooperation under Nash bargaining, $\nu:=0.09, \sigma:=0.15, \mu:=\nu+\sigma^2/2, r:=0.03$.}
\label{cooperation}
\end{center}
\end{figure}

Remembering that we have normalized the gambler's fortune to $\$1$, the broker faces the demand curve
\begin{equation}
q\left(r_L\right)=b\left(r_L\right)-1=\left(\frac{\mu}{\sigma^2}-1\right)-\frac{1}{\sigma^2}\times r_L.
\end{equation}The corresponding inverse demand (or marginal value) curve is
$r_L(q)=\left(\mu-\sigma^2\right)-\sigma^2q$, and the marginal revenue curve is $\text{MR}(q)=\left(\mu-\sigma^2\right)-2\sigma^2q$.  The instantaneous price elasticity of demand for margin loans is
\begin{equation}
\epsilon^d(q)=-\frac{r_L}{q}\times\frac{dq}{dr_L}=\frac{\mu-\sigma^2}{\sigma^2q}-1.
\end{equation}Equating the broker's marginal revenue to the marginal cost $r$ of funding, we obtain the monopoly quantity
\begin{equation}
q_M=\frac{\mu-\sigma^2-r}{2\sigma^2}.
\end{equation}
Substituting this quantity into the inverse demand curve, we get the familiar monopoly midpoint pricing rule:
\begin{equation}
r_M=\frac{\mu-\sigma^2+r}{2}.
\end{equation}That is, the monopoly price of margin loans is equal to the average of the choke point $\mu-\sigma^2$ and the broker call rate $r$.  Thus, under the monopoly threat point, the continuous-time Kelly gambler will bet the fraction
\begin{equation}
b_M=q_M+1=\frac{\mu+\sigma^2-r}{2\sigma^2}
\end{equation}of wealth on the risk asset over $\left[t,t+dt\right]$.  Given its net interest margin of $r_M-r=\left(\mu-\sigma^2-r\right)/2$, the broker's instantaneous rate of intermediation profit per dollar of client equity is
\begin{equation}
\pi_M=\left(r_M-r\right)q_M=\left(\frac{\mu-\sigma-r}{2\sigma}\right)^2,
\end{equation}and the gambler's asymptotic capital growth rate under the monopoly market structure is
\begin{equation}
\Gamma_M=\Gamma\left(b_M,r_M\right)=\frac{\mu-\sigma^2+r}{2}+\frac{1}{8}\left(\frac{\mu+\sigma^2-r}{\sigma}\right)^2.
\end{equation}Consumer surplus flows to the Kelly gambler at a continuous rate of 
\begin{equation}
\text{CS}=\frac{1}{8}\left(\frac{\mu-\sigma^2-r}{\sigma}\right)^2
\end{equation}per unit time.  From the symmetry of the monopoly midpoint, the deadweight loss per unit time is equal to the consumer surplus:
\begin{equation}
\text{DWL}=\text{CS}=\frac{1}{8}\left(\frac{\mu-\sigma^2-r}{\sigma}\right)^2.
\end{equation}
\begin{example}
For the parameters $\nu:=0.09, \sigma:=0.15, \mu:=\nu+\sigma^2/2, r:=0.03$, the monopoly price of margin debt is $r_M=5.44\%$, and the gambler correspondingly would demand $q_M=1.083$ dollars of margin loans per dollar of his account equity, for a total bet of $b_M=2.0833$.  The gambler thereby achieves an asymptotic capital growth rate of $\overline{\Gamma}=10.32\%$ and the broker earns intermediation profits at annual rate of $\overline{\pi}=2.64\%$ of client equity.  This monopoly behavior is illustrated in Figure \ref{monopoly}.

\begin{figure}[t]
\begin{center}
\includegraphics[height=250px]{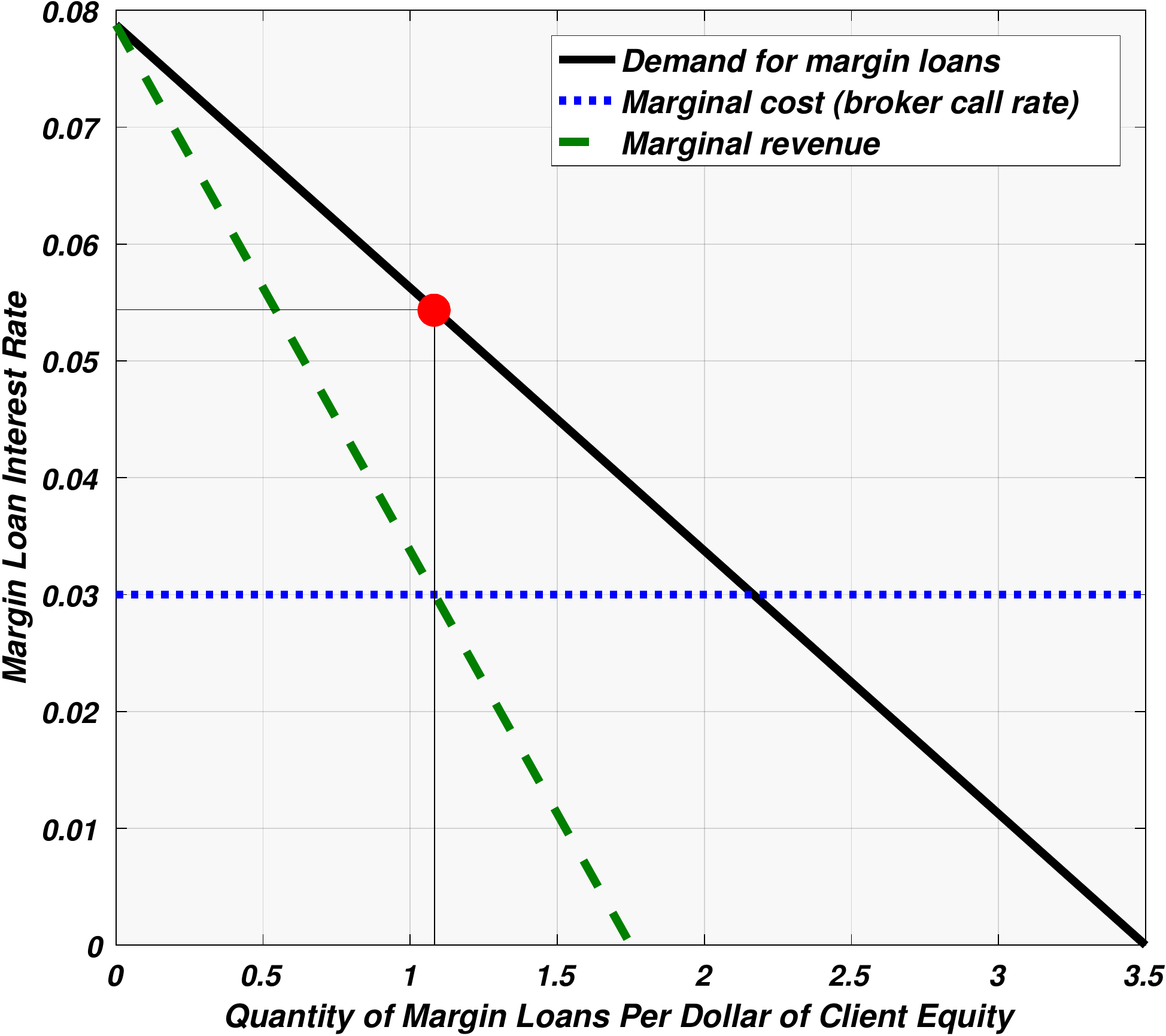}
\caption{\sc Kelly gambler's instantaneous demand for margin loans per dollar of account equity, $\nu:=0.09, \sigma:=0.15, \mu:=\nu+\sigma^2/2, r:=0.03$.}
\label{monopoly}
\end{center}
\end{figure}

\par
Under cooperation (relative to this threat point), the gambler ups his bet to $b^*=3.17$ and the broker lowers his interest rate to $r_L^*=4.52\%$.  This raises the gambler's growth rate to $10.98\%$ and it raises the broker's profit rate to $3.3\%$.  Thus, the annual rate of deadweight loss (in the amount of $1.32\%$ of account equity) has been converted to surplus value and shared equally between the counterparties:  $\Gamma-\overline{\Gamma}=\pi-\overline{\pi}=0.66\%$.  This cooperative behavior is illustrated in Figure (\ref{cooperation}).
\end{example}

\section{Several Risk Assets.}
In this section, we proceed to extend our main techniques and results to the general stock market with $n$ correlated risk assets ($i:=1,2,...,n$) in geometric Brownian motion.  Let $S_{it}$ denote the price of stock $i$ at time $t$, where
\begin{equation}
\frac{dS_{it}}{S_{it}}:=\mu_i\,dt+\sigma_i\,dW_{it}.
\end{equation}$\mu_i$ is the drift of stock $i$, $\sigma_i$ is its volatility, and the $\left(W_{it}\right)_{i=1}^n$ are correlated unit Brownian motions.  On that score, we let $\rho_{ij}:=\text{Corr}\left(dW_{it},dW_{jt}\right)$, and we let 
\begin{equation}
\sigma_{ij}:=\rho_{ij}\sigma_i\sigma_j=\text{Cov}\left(\frac{dS_{it}}{S_{it}},\frac{dS_{jt}}{S_{jt}}\right)\bigg/dt
\end{equation}
denote the covariance of instantaneous returns per unit time.  In what follows, we will let $\mu:=\left(\mu_1,...,\mu_n\right)'$ denote the drift vector of the stock market, and we will assume that the covariance matrix $\Sigma:=\left[\sigma_{ij}\right]_{n\times n}$ is invertible.  In this generality, a continuously-rebalanced portfolio (or fixed-fraction betting scheme) is defined by a vector $b:=\left(b_1,...,b_n\right)'\in\mathbb{R}^n$ of portfolio weights, where the intention is to continuously execute rebalancing trades so as to maintain the fixed fraction $b_i$ of wealth in stock $i$ at all times.  As usual, we let $V_t(b)$ denote the gambler's fortune at time $t$;  the instantaneous quantity of margin loans demanded therefore amounts to 
\begin{equation}
q=\left(\sum\limits_{i=1}^nb_i-1\right)V_t(b)=\left(\textbf{1}'b-1\right)V_t(b),
\end{equation}or just $q=\textbf{1}'b-1$ per dollar of client equity, where $\textbf{1}:=\left(1,...,1\right)'$ is an $n\times1$ vector of ones.  The broker's instantaneous rate of intermediation profit is now $\pi=\left(\textbf{1}'b-1\right)\left(r_L-r\right)$ per dollar of client equity.  The evolution of the gambler's fortune is now governed by the stochastic differential equation
\begin{equation}
\frac{dV_t(b)}{V_t(b)}:=\sum_{i=1}^nb_i\frac{dS_{it}}{S_{it}}-\left(\textbf{1}'b-1\right)r_L\,dt=\left[r_L+\left(\mu-r_L\textbf{1}\right)'b\right]dt+\sum_{i=1}^nb_i\sigma_i\,dW_{it}.
\end{equation}Applying It\^{o}'s Lemma for several diffusion processes (cf. with Paul Wilmott 1998), we obtain the fact that
\begin{equation}
d\left(\log V_t(b)\right)=\left[r_L+\left(\mu-r_L\textbf{1}\right)'b-\frac{1}{2}b'\Sigma b\right]\,dt+\sum_{i=1}^nb_i\sigma_i\,dW_{it},
\end{equation}which, upon integration, yields
\begin{equation}
V_t(b)=V_0\times\exp\left\{\left[r_L+\left(\mu-r_L\textbf{1}\right)'b-\frac{1}{2}b'\Sigma b\right]t+\sum_{i=1}^nb_i\sigma_iW_{it}\right\}.
\end{equation}In this connection, the gambler's continuously-compounded asymptotic capital growth rate is now
\begin{equation}
\boxed{\Gamma\left(b,r_L\right):=r_L+\left(\mu-r_L\textbf{1}\right)'b-\frac{1}{2}b'\Sigma b=\lim_{t\to\infty}\frac{\log\left[V_t(b)/V_0\right]}{t}=\frac{\mathbb{E}\left[d\left(\log V_t(b)\right)\right]}{dt}.}
\end{equation}Proceeding as before, the correct behavior is obtained by optimizing the Nash product $N\left(b,r_L\right)$:
\begin{equation}
\left(b^*,r_L^*\right):=\argmax_{\left\{\textbf{1}'b\geq1,\,\,r_L\geq r\right\}}\,\,\left(\pi-\overline{\pi}\right)\left(\Gamma-\overline{\Gamma}\right).
\end{equation}Using the product rule to calculate $\partial N/\partial r_L$, we obtain
\begin{equation}
\left(1-\textbf{1}'b\right)\left(\pi-\overline{\pi}\right)+\left(\Gamma-\overline{\Gamma}\right)\left(\textbf{1}'b-1\right)=0.
\end{equation}Assuming interiority $(\textbf{1}'b>1)$, we cancel the common factor $q=\textbf{1}'b-1$ and again obtain the egalitarian condition $\Gamma-\overline{\Gamma}=\pi-\overline{\pi}$.  Using the product rule to calculate the gradient $\nabla_bN$ of the Nash product with respect to $b$, we get the (vector) condition
\begin{equation}
\left(\pi-\overline{\pi}\right)\left(\mu-r_L\textbf{1}-\Sigma b\right)+\left(\overline{\Gamma}-\Gamma\right)\left(r_L-r\right)\textbf{1}=\textbf{0}.
\end{equation}Assuming that there are gains to be had from cooperation (meaning that $\pi-\overline{\pi}$ and $\Gamma-\overline{\Gamma}$ are both positive numbers), we cancel this common factor and simplify to obtain
\begin{equation}
\boxed{b^*=\Sigma^{-1}\left(\mu-r\textbf{1}\right).}
\end{equation}As expected, this is precisely the behavior of a continuous time Kelly gambler (cf. with David Luenberger 1998) who is permitted to borrow cash at the broker's call rate, $r$.  Using this fact in conjunction with the egalitarian condition and the defining expressions for $\Gamma$ and $\pi$ one calculates the negotiated interest rate to be
\begin{equation}
\boxed{r_L^*=\frac{r}{2}+\frac{\mu'\Sigma^{-1}\mu-r^2\cdot\textbf{1}'\Sigma^{-1}\textbf{1}-2\left(\overline{\Gamma}-\overline{\pi}\right)}{4\left[\textbf{1}'\Sigma^{-1}\left(\mu-r\textbf{1}\right)-1\right]},}
\end{equation}which is in perfect accord with (\ref{negotiated}).
\par
Taking our cue from the univariate case, we can derive the (linear) efficient $\Gamma-\pi$ frontier in just the same way.  For a given rate $\pi$ of intermediation profit, we solve for $r_L$ and obtain
\begin{equation}
r_L=r+\frac{\pi}{\textbf{1}'b-1}.
\end{equation}Substituting this expression into the definition of $\Gamma$, one has, after simplification,
\begin{equation}\label{maxout}
\Gamma=r+\left(\mu-r\textbf{1}\right)'b-\frac{1}{2}b'\Sigma b-\pi.
\end{equation}Maximizing $b$ out of this expression, we obtain $b^*=\Sigma^{-1}\left(\mu-r\textbf{1}\right)$;  the general equation of the efficient frontier is
\begin{equation}\label{efficient}
\boxed{\Gamma=r+\frac{1}{2}\left(\mu-r\textbf{1}\right)'\Sigma^{-1}\left(\mu-r\textbf{1}\right)-\pi.}
\end{equation}The final utility vector $\left(\Gamma^*,\pi^*\right)$ that obtains from Nash bargaining therefore lies at the intersection of the efficient frontier (\ref{efficient}) and the line $\Gamma=\overline{\Gamma}-\overline{\pi}+\pi$ which expresses the egalitarian division of surplus value.  Solving these simultaneous equations, we get
\begin{equation}
\boxed{\Gamma^*=\frac{r+\overline{\Gamma}-\overline{\pi}}{2}+\frac{1}{4}\left(\mu-r\textbf{1}\right)'\Sigma^{-1}\left(\mu-r\textbf{1}\right)}
\end{equation}and
\begin{equation}
\boxed{\pi^*=\frac{r-\left(\overline{\Gamma}-\overline{\pi}\right)}{2}+\frac{1}{4}\left(\mu-r\textbf{1}\right)'\Sigma^{-1}\left(\mu-r\textbf{1}\right).}
\end{equation}
\subsection{Monopoly Disagreement Point.}
If negotiations break down and the broker simply posts a price $r_L$, then the gambler will react with the corresponding Kelly rule, namely, $b(r_L)=\Sigma^{-1}\left(\mu-r_L\textbf{1}\right).$  Thus, the broker faces the instantaneous demand curve
\begin{equation}
q(r_L)=\textbf{1}'b(r_L)-1=\left(\textbf{1}'\Sigma^{-1}\mu-1\right)-\left(\textbf{1}'\Sigma^{-1}\textbf{1}\right)r_L.
\end{equation}The instantaneous elasticity of demand for margin loans is therefore given by
\begin{equation}
\epsilon^d(q)=\frac{\textbf{1}'\Sigma^{-1}\mu-1}{q}-1.
\end{equation}The inverse demand (or marginal value) curve is
\begin{equation}
r_L=\text{MV}(q)=\frac{\textbf{1}'\Sigma^{-1}\mu-1}{\textbf{1}'\Sigma^{-1}\textbf{1}}-\frac{1}{\textbf{1}'\Sigma^{-1}\textbf{1}}\times q,
\end{equation}which induces the marginal revenue curve
\begin{equation}
\text{MR}(q)=\frac{\textbf{1}'\Sigma^{-1}\mu-1}{\textbf{1}'\Sigma^{-1}\textbf{1}}-\frac{2}{\textbf{1}'\Sigma^{-1}\textbf{1}}\times q.
\end{equation}Intersecting marginal revenue with marginal cost (which is $\text{MC}(q):\equiv r$), we get the monopoly quantity
\begin{equation}
\boxed{q_M=\frac{\textbf{1}'\Sigma^{-1}\mu-1-r\cdot\textbf{1}'\Sigma^{-1}\textbf{1}}{2}.}
\end{equation}Reading off the inverse demand curve, the monopoly interest rate is now
\begin{equation}
\boxed{r_M=\frac{1}{2}\left(\frac{\textbf{1}'\Sigma^{-1}\mu-1}{\textbf{1}'\Sigma^{-1}\textbf{1}}+r\right).}
\end{equation}Exact formulas for all remaining quantities of interest, like the consumer surplus, the deadweight loss, and the profit and growth rates that would obtain under the monopoly threat point, all follow in the obvious way from the monopoly price and quantity given above, just as they did in the univariate case.
\subsection{General Solution for CRRA Utility.}
To close the paper, we indicate briefly how our results can be (easily) extended for the benefit of arbitrary CRRA investors, e.g. those whose preferences over terminal wealth can be represented by the isoelastic utility function $u(x):=x^{1-\gamma}$, where $\gamma>0$ is the agent's (constant) coefficient of relative risk aversion. 
\par
Applying It\^{o}'s Lemma, one can calculate that (for $\gamma\neq1$) the continuously rebalanced portfolio $b:=\left(b_1,...,b_n\right)'\in\mathbb{R}^n$ generates the the following law of motion for the investor's utility $V_t(b)^{1-\gamma}$:
\begin{equation}
\left(1-\gamma\right)^{-1}\frac{d\left(V_t(b)^{1-\gamma}\right)}{V_t(b)^{1-\gamma}}=\left[r_L+(\mu-r_L\textbf{1})'b-\frac{\gamma}{2}b'\Sigma b\right]dt+\sum_{i=1}^nb_i\sigma_i\,dW_{it}.
\end{equation}Thus, the ``growth rate'' $\Gamma$ now takes the form
\begin{equation}
\Gamma\left(b,r_L\right):=r_L+\left(\mu-r_L\textbf{1}\right)'b-\frac{\gamma}{2}b'\Sigma b=\left(1-\gamma\right)^{-1}\frac{\mathbb{E}_t\left[d\left(V_t(b)^{1-\gamma}\right)\right]}{V_t(b)^{1-\gamma}}\bigg/dt,
\end{equation}e.g. $\Gamma\left(b,r_L\right)$ is directly proportional to the expected percent change in (power) utility per unit time over the differential time step $\left[t,t+dt\right]$. 
\par
\textit{De facto}, then, the only formal change in the model is that the covariance matrix $\Sigma$ has been replaced by $\gamma\Sigma$;  in the univariate case this means e.g. that the variance $\sigma^2$ must everywhere be replaced by $\gamma\sigma^2$.  Thus, the investor negotiates to use the continuously-rebalanced portfolio $b^*=\left(1/\gamma\right)\Sigma^{-1}\left(\mu-r\textbf{1}\right)$, where $r$ is the broker's cost of funds.  The equilibrium interest rate now amounts to
\begin{equation}
r_L^*=\frac{r}{2}+\frac{\mu'\Sigma^{-1}\mu-r^2\cdot\textbf{1}'\Sigma^{-1}\textbf{1}-2\gamma\left(\overline{\Gamma}-\overline{\pi}\right)}{4\left[\textbf{1}'\Sigma^{-1}\left(\mu-r\textbf{1}\right)-\gamma\right]}.
\end{equation}The efficient frontier in the $\left(\pi,\Gamma\right)$-plane is now given by the equation
\begin{equation}
\Gamma=r+\frac{1}{2\gamma}\left(\mu-r\textbf{1}\right)'\Sigma^{-1}\left(\mu-r\textbf{1}\right)-\pi,
\end{equation}so that phenomenon of transferable utility is preserved for general CRRA investors.  
\par
In parting, we note that the stock market parameters $\left(\mu,\Sigma\right)$ must be assumed to be sufficiently favorable (and the penalty parameters $r,\gamma$ must be sufficiently low) that the investor is at least willing to borrow money at the broker's own cost of funds;  this means that the deep parameters $\left(\mu,\Sigma,r,\gamma\right)$ must satisfy the inequality
\begin{equation}
\textbf{1}'\Sigma^{-1}\left(\mu-r\textbf{1}\right)>\gamma.
\end{equation}

\section{Summary and Conclusions.}
This paper studied negotiated margin loan contracts between continuous time Kelly gamblers (more generally, CRRA investors) and the brokers who lend them cash to make large bets on high-growth assets or portfolios.  On account of the continuous sample path of asset prices, the broker bears no risk of default, for it can in principle liquidate the client's assets at the very instant his account equity is equal to zero.  In the Black-Scholes market consisting of several correlated stocks in geometric Brownian motion, the gambler's fortune itself follows a geometric Brownian motion, and therefore remains positive until the hereafter.  A margin loan quantity of $q:=\textbf{1}'b-1$ per dollar of client equity is made for the duration of the differential time step $\left[t,t+dt\right]$, after which interest is debited and the size of the loan is readjusted on account of the observed fluctuation $dV_t(b)$ of the gambler's bankroll.
\par
All potential cooperation between these counterparties rests on the fact that the market admits assets (or continuously-rebalanced portfolios of said assets) whose asymptotic growth rate is significantly higher than the broker's call rate, $r$.  The broker, who charges margin interest at a continuous rate of $r_L>r$, seeks an arrangement whereby his instantaneous rate of intermediation profit, namely $\pi:=\left(\textbf{1}'b-1\right)\left(r_L-r\right)$ per dollar of client equity, is as high as possible.  The Kelly gambler, who is notoriously far sighted (cf. with MacLean, Thorp, and Ziemba 2011), is willing to stomach any level of volatility, value-at-risk, or maximum drawdown in exchange for the highest possible asymptotic capital growth rate, here denoted $\Gamma:=r_L+\left(\mu-r_L\textbf{1}\right)'b-b'\Sigma b/2$.  In an evolutionary sense, the broker must treat his Kelly gamblers with kid gloves, for they will hold asymptotically $100\%$ of the equity on deposit, and (on account of their fixed leverage ratio) they will shoulder $100\%$ of all margin debt in the limit.
\par
If the broker and the Kelly gambler fail to come to terms (that would simultaneously specify the interest rate, the portfolio, and the quantity of margin loans), then there are two obvious ways the disagreement could play out.  In the worst scenario, the Kelly gambler borrows no cash at all (or the broker doesn't lend him any);  the broker's intermediation profit is zero and the client makes do with optimizing his growth rate over the set of unlevered continuously-rebalanced portfolios (e.g. he is constrained by the condition $\textbf{1}'b=1$).  The second possibility is the one that seems to actually obtain in this world:  the broker just posts a monopoly price $r_M$, and the gambler then demands the corresponding monopoly quantity $q_M$ of margin loans as dictated by his instantaneous demand curve.
\par
Following Nash's (1950) theory of axiomatic bargaining, we maximized the Nash product $N:=\left(\pi-\overline{\pi}\right)\times\left(\Gamma-\overline{\Gamma}\right)$ over the profit-growth plane of points $\left(\pi,\Gamma\right)\geq\left(\overline{\pi},\overline{\Gamma}\right)$; the factors of this product are the respective surplus values extracted by the counterparies from  cooperation as opposed to disagreement, which only yields meager  levels $\left(\overline{\pi},\overline{\Gamma}\right)$ of utility.  We found that, regardless of the threat point $\left(\overline{\pi},\overline{\Gamma}\right)$, the gambler will negotiate to bet as if he himself could borrow money at the broker's (low) cost of funds.  On account of the egalitarian (first-order) condition $\Gamma-\overline{\Gamma}=\pi-\overline{\pi}$, the surplus value from cooperation will in any event be divided evenly between the counterparties.  We used this fact to derive exact formulas for the negotiated interest rate $r_L^*$ and the final utilities $\left(\pi^*,\Gamma^*\right)$ that obtain from cooperation.  Finally, we derived an expression for the efficient frontier in the profit-growth plane:  it is a straight line (whose slope is $-1$), meaning that our particular bargaining problem enjoys the special property of transferable utility (cf. with Roger Myerson 1997).  Geometrically, the negotiated outcome $\left(\pi^*,\Gamma^*\right)$ is the result of intersecting the efficient frontier with a $45$\textdegree\, line emanating from the threat point $\left(\overline{\pi},\overline{\Gamma}\right)$.
\par
\textit{\sc Northern Illinois University}

\newpage
\subsubsection*{\sc\underline{Acknowledgments}}
I thank Omri Tal, Nero Tulip, and several anonymous referees for helpful comments, discussions, and suggestions that improved the paper.
\subsubsection*{\sc\underline{Funding}}
This work was supported by the Open Access Publishing Fund at Northern Illinois University, which is administered by the University Libraries and funded by the College of Liberal Arts and Sciences, Research and Innovation Partnerships (RIPS), and the University Libraries.
\subsubsection*{\sc\underline{Availability of Data and Materials}}
N/A.
\subsubsection*{\sc\underline{Authors' Contributions}}
This paper is solely the work of the author.
\subsubsection*{\sc\underline{Competing Interests}}
The author declares that he has no conflicts of interest.
\subsubsection*{\sc\underline{Author Details}}
Department of Economics, School of Public and Global Affairs, College of Liberal Arts and Sciences, Northern Illinois University, DeKalb IL 60115.

\end{document}